# WATER-CHANNEL STUDY OF FLOW AND TURBULENCE PAST A TWO-DIMENSIONAL ARRAY OF OBSTACLES

## Annalisa Di Bernardino[1], Paolo Monti[1], Giovanni Leuzzi[1], Giorgio Querzoli[2]

[1]DICEA, Università di Roma "La Sapienza". Via Eudossiana 18 - 00184, Roma. Italy.

[2]Dipartimento di Ingegneria del Territorio, Università di Cagliari, Via Marengo 3 – 09123, Cagliari. Italy.

**Abstract**     A neutral boundary layer was generated in the laboratory to analyze the mean velocity field and the turbulence field within and above an array of two-dimensional obstacles simulating an urban canopy. Different geometrical configurations were considered in order to investigate the main characteristics of the flow as a function of the aspect ratio ($AR$) of the canopy. To this end, a summary of the two-dimensional fields of the fundamental turbulence parameters is given for $AR$ ranging from 1 to 2. The results show that the flow field depends strongly on $AR$ only within the canyon, while the outer flow seems to be less sensitive to this parameter. This is not true for the vertical momentum flux, which is one of the parameters most affected by $AR$, both within and outside the canyon. The experiments also indicate that, when $AR \lesssim 1.5$ (i.e. the skimming flow regime), the roughness sub-layer extends up to a height equal to 1.25 times the height of the obstacles ($H$), surmounted by an inertial sub-layer that extends up to 2.7 $H$. In contrast, for $AR > 1.5$ (i.e. the wake-interference regime) the inertial sub-layer is not present. This has significant implications when using similarity laws for deriving wind and turbulence profiles in canopy flows. Furthermore, two estimations of the viscous dissipation rate of turbulent kinetic energy of the flow are given. The first one is based on the fluctuating strain rate tensor, while the second is related to the mean strain rate tensor. It is shown that the two expressions give similar results, but the former is more complicated, suggesting that the latter might be used in numerical models with a certain degree of reliability. Finally, the data presented can also be used as a dataset for the validation of numerical models.

Corresponding author: Paolo Monti. DICEA, Università di Roma "La Sapienza". Via Eudossiana 18 – 00184, Roma. Italy. E-mail: paolo.monti@uniroma1.it

**1 Introduction**

The rapid growth of population experienced in large cities over the last few decades has led to the increase of air pollution in urban areas, causing degradation of environmental quality and human comfort. Much effort has therefore been made into the analysis of flow and dispersion within urban environments (Fernando et al. 2001). Despite the fact that significant progress has been made on the understanding of urban fluid mechanics, a variety of issues still remains unresolved (Fernando, 2010).

In the literature, special attention is paid to the street canyon, assumed as an archetype for more complex and realistic urban fabrics. Hussain and Lee (1980) found that one of the most important parameters to be considered is the aspect ratio $AR = W / H$, i.e. the ratio of the spacing between buildings, $W$, to the height of the buildings, $H$. Based on past studies conducted in wind tunnels and water channels, Oke (1987) summarized the nature of the flow in urban canopies in terms of $AR$ in the case of neutral conditions. He defined three kinds of flow regimes: the skimming flow ($AR \lesssim 1.5$), in which only a single vortex develops within the street canyon; the wake-interference flow ($1.5 \lesssim AR \lesssim 2.5$), which allows the development of two counter-rotating vortexes; and the isolated obstacle regime ($AR \gtrsim 2.5$), where the flow strictly resembles that observed for the isolated building case.

Several computational fluid dynamics simulations have recently been conducted to examine the mean and turbulent characteristics of the flow over arrays of 3D buildings. Kanda et al. (2004) performed a large eddy simulation (LES) to study the well-organized turbulence structures that form above the building canopy, while Xie and Castro (2006) examined the turbulent flow over staggered wall-mounted cubes. Cui et al. (2004), Liu et al. (2004) and Gowardhan et al. (2007) employed a similar approach to analyze the turbulence within the canyon, while Hang et al. (2012) investigated numerically the influence of the building-height variability on city breathability. Recently, LES models have also been used to investigate turbulent flows in densely built-up urban areas (Park et al. 2013).

A number of experiments have also been conducted in the laboratory with the aim of reproducing urban canopies (see, for example, the comprehensive review provided by Ahmad et al. 2005). Uehara et al. (2000) used the wind tunnel to study the turbulence characteristics in a regular array of 3D buildings, and focused their attention on the effects of atmospheric stability on the flow within a street canyon. Cheng and Castro (2002), on the basis of a series of experiments conducted in the wind tunnel, examined in detail the strong three-dimensionality of the turbulence in the roughness sub-layer (RSL), i.e. the region above the canopy where the flow is influenced by the individual roughness elements. They also estimated the depth of the inertial sub-layer (ISL, i.e. the region above the RSL where the turbulent fluxes are nearly constant with height and the usual rough-wall logarithmic velocity law applies) for each building configuration. Princevac et al. (2010) analyzed the flow field in a water-channel on vertical and horizontal planes in correspondence with a 3D building array, with the focus on the lateral channelling. Water-channel studies were also conducted by Huq and Franzese (2013), who made turbulence and scalar concentration measurements at different heights within an array of buildings for different $AR$ values.

In the last few decades, several field campaigns were also conducted with the aim of delineating urban flows and pollutant dispersion in cities. For example, meteorological and dispersion datasets at near full-scale were built during the Mock Urban Setting Test (MUST) for the development and validation of urban toxic hazard assessment models (Biltoft, 2001). In the Joint Urban 2003 experiment, a multi-group team studied the Oklahoma City urban boundary layer with a high density of instrumentation, while the Basel UrBan Boundary Layer Experiment (BUBBLE) allowed detailed investigation of the boundary-layer structure above the City of Basel, Switzerland (Rotach et al. 2005).

Laboratory scale studies have also been conducted to analyze flow and dispersion in arrays of 2D canyons. For example, Baik et al. (2000) focused their attention on the influence of the aspect ratio on the turbulent field, while Kastner-Klein et al. (2001) quantified the effects of vehicular traffic on the airflow in the canyon. Salizzoni et al. (2011) examined the turbulent transfer generated by the shear layer above the canyons and found that this transfer process cannot be expressed in a non-dimensional form based on a single velocity scale. Soulhac et al. (2008) numerically simulated 2D street-canyon flows and proposed a theoretical model to describe the flow along a 2D street canyon for any external wind direction. Useful insight into the physics of this problem was provided also by the numerical studies reported in Casonato and Gallerano (1990), Kim and Baik (1999, 2001), Jeong and Andrews (2002), Lien et al. (2004) and Li et al. (2010).

Here a laboratory investigation of the neutrally-stratified boundary layer that forms above and within a two-dimensional array of buildings is conducted in a water-channel experiment. Although considerable progress has recently been made in gaining an exhaustive knowledge of street-canyon flows, further work is needed to reach a more complete quantitative description (Pelliccioni et al. 2014). The present work is therefore motivated by the belief that laboratory experiments can provide useful information on flow and dispersion within urban canopies that have general applicability for mesoscale studies of the urban heat island (see, for example, Salamanca et al. 2010; Cantelli et al. 2014; Luhar et al. 2014) as well as for numerical predictions of pollutant concentration in urban canopies (Leuzzi et al. 2012).

Our goal is to examine the turbulence characteristics in urban street canyons as well as the processes by which the flow within the canopy layer exchanges energy and momentum with the overlaying fluid layer. In Sect. 2 we describe the experimental set-up used for the experiments, and in Sect. 3 we present the results, while conclusions are given in Sect. 4.

## 2 Experimental set-up and measurement technique

The facility is located at the Hydraulics Laboratory of the University of Rome - La Sapienza, Italy. A closed-loop water-channel is used for the experiments (Fig. 1); the channel is 0.35 m high, 0.25 m wide and 7.40 m long, and the flume is fed by a constant head reservoir. In the first part of the channel, three honeycombs minimize secondary flows and other unwanted effects associated with the inlet system. A floodgate positioned at the end of the channel permits the regulation of the water depth and, therefore, of the water velocity. For all the experiments, the water depth is set to $h$ = 0.16 m. The channel bottom is covered by small pebbles - average size 0.005 m - in order to increase the surface roughness. The test section is positioned 5 m downstream of the inlet, where the boundary layer is fully developed.

The investigated urban canopy consists of a 2D array of obstacles, with a series of parallelepipeds of square section $B = H = 0.02$ m and length $L = 0.25$ m fixed to the channel bottom. During the tests the distance between buildings, $W$, is varied from 0.02 m up to 0.04 m, and correspondingly, the aspect ratio ($AR$) of the canopy ranges from 1 up to 2. Cases $AR = 1$ and 2 are examined in detail, whereas the intermediate cases $AR = 1.5$ and 1.75 will be considered occasionally. A series of 20 buildings is placed upstream of the investigated area in order to obtain a fully-developed flow.

Flow velocity is measured by image analysis, whereby the working fluid is seeded with non-buoyant particles, $2\times10^{-5}$ m in diameter, and a high-speed camera (CMOS Camera with a resolution of 1280 × 1024 pixels) acquires videos at 250 frames per second for a duration of 40 s. A thin laser light sheet (wavelength: 532 nm; depth: 0.002 m) illuminates the test section.

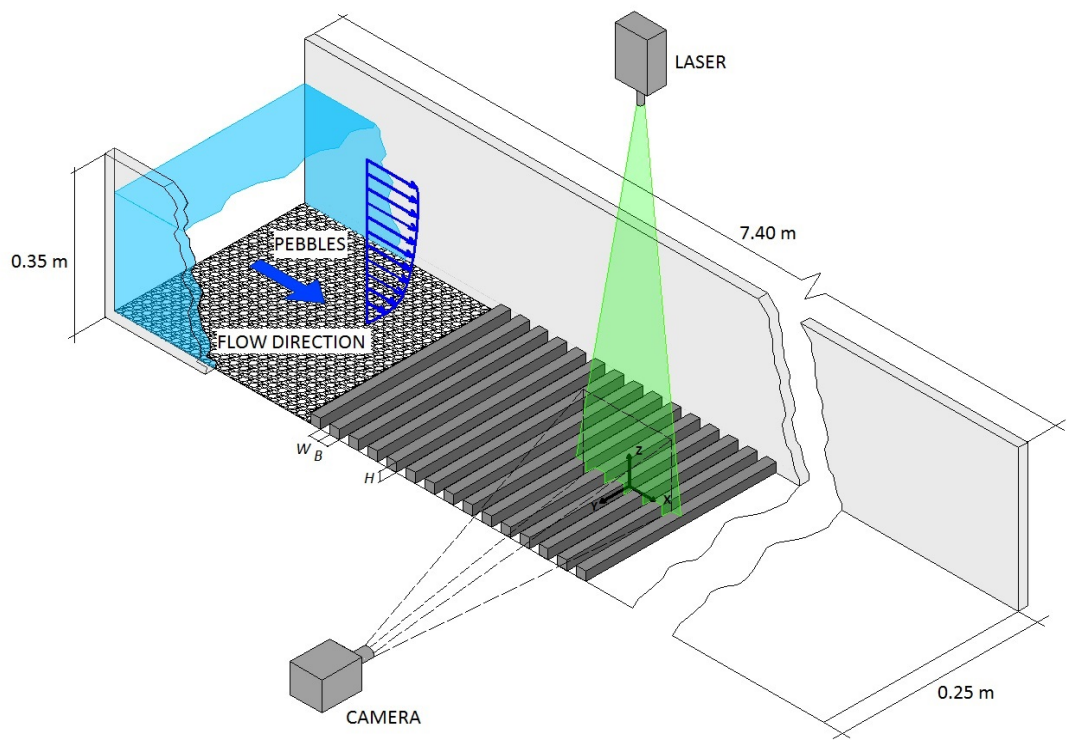

**Fig. 1** Scheme of the modelled urban canopy ($AR = 1$ is represented). $H$ indicates the building height while $B = H$ is its length. $W$ is the distance between two successive buildings (i.e. the street width). The $x$-axis refers to the channel axis, while the $z$-axis is parallel to the vertical

The images taken by the high-speed camera are analyzed with a feature tracking algorithm that recognises particle trajectories. Velocities are deduced from particle displacements between successive frames and interpolated on a regular grid by Gaussian averaging. The resulting spatial resolution is 1 mm. This method has already been used in several studies (e.g. Cenedese et al. 2005; Fortini et al. 2013); details can be found in Miozzi et al. (2008).

Since the upper surface of the buildings spreads light, preventing a successful particle recognition and, consequently, reliable velocity measurements, a 0.002-m thick layer above the top of the buildings is excluded from the following analysis.

The framed area is rectangular, lying in the vertical mid-plane of the channel, 0.099 m long and 0.072 m high. We define a reference frame with the $x$-axis aligned with the streamwise velocity, the $z$-axis vertical and the $y$-axis in the spanwise direction. The origin is on the mid-plane, $x$ is measured from the centre of the investigated canyon and $z$ from the ground upwards. The Reynolds number of the flow is $Re = (Uh/\nu) \cong 44,000$, where $U = 0.27$ m s$^{-1}$ is the stream free velocity and $\nu = 10^{-6}$ m$^2$ s$^{-1}$ is the kinematic viscosity of water. As a consequence, $Re$ is well-above the critical value in order that both the simulated large-scale structures and the mean flow can be considered to be independent of $Re$ (Snyder, 1981).

## 3 Results and discussion

### 3.1 Mean velocity and variance

Figures 2a and 2b report a vector representation of the mean velocity referred to $AR = 1$ and 2, respectively. The values are non-dimensionalized by the stream free velocity, directed rightwards. For $AR = 1$ the flow pattern conforms to the classical configuration of the skimming flow, i.e. a current above the canopy nearly parallel to the $x$-direction and a main vortex within the canyon, the latter characterized by lower speeds. The vortex centre is slightly shifted downstream and towards the top of the canyon, implying higher velocity in the descending flow near the windward building with respect to the ascending flow close to the leeward building. In agreement with the LES results of Li et al. (2010), a small, counter-rotating vortex forms at the bottom of the windward building. In contrast, for $AR = 2$ (wake interference flow, Fig. 2b), the main vortex is significantly shifted downstream and a well-defined counter-rotating vortex forms near the leeward building (see the LES results of Liu et al. 2004 and Brevis et al. 2014).

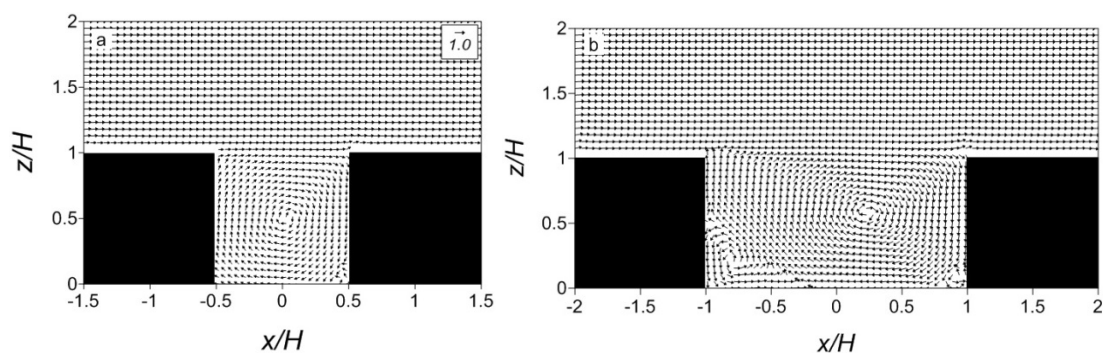

**Fig. 2** Non-dimensional mean velocity vectors for $AR = 1$ (a) and $AR = 2$ (b). Velocity components are expressed as $\bar{u}/U$ and $\bar{w}/U$

According to numerical evidence reported in the literature (see, for example, Li et al. 2010), the variance of the non-dimensional horizontal velocity component, $\overline{u'^2}/U^2$, (here primes are fluctuations around the mean) assumes lower values inside the canyon (nearly one order of magnitude) irrespective of $AR$ (Figs. 3a and b). In contrast, the non-dimensional vertical velocity variance, $\overline{w'^2}/U^2$, shows large values within a tongue-like feature protruding from the outer flow, near the windward wall when $AR = 1$ (Fig. 4a). That feature is larger for $AR = 2$ (Fig. 4b), where in the right-half of the canyon $\overline{w'^2}/U^2$ is of the same order as in the outer flow. It should be underlined that $\overline{w'^2}/U^2$ reaches a local minimum close to the rooftop, in agreement with Li et al. (2010), while $\overline{u'^2}/U^2$ reaches its maximum there. This agrees with the results of Salizzoni et al. (2011) obtained in the wind tunnel for $AR = 1$.

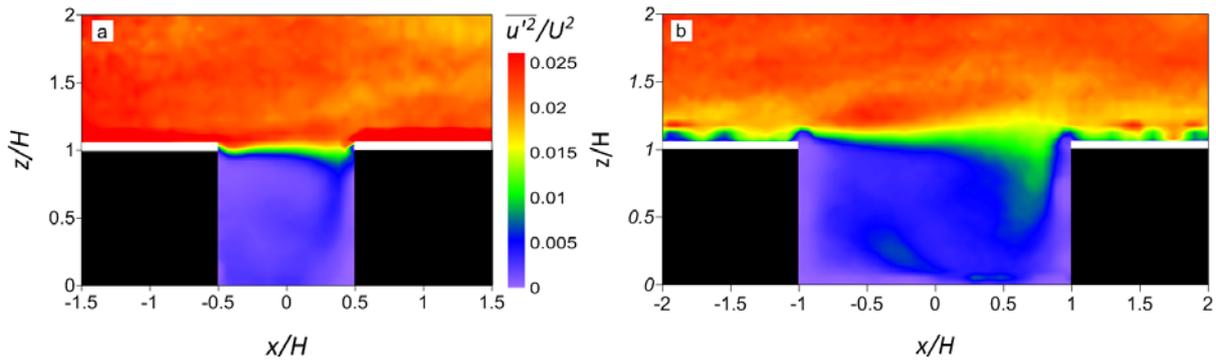

**Fig. 3** Non-dimensional horizontal velocity variance $\overline{u'^2}/U^2$ maps for $AR = 1$ (a) and $AR = 2$ (b).

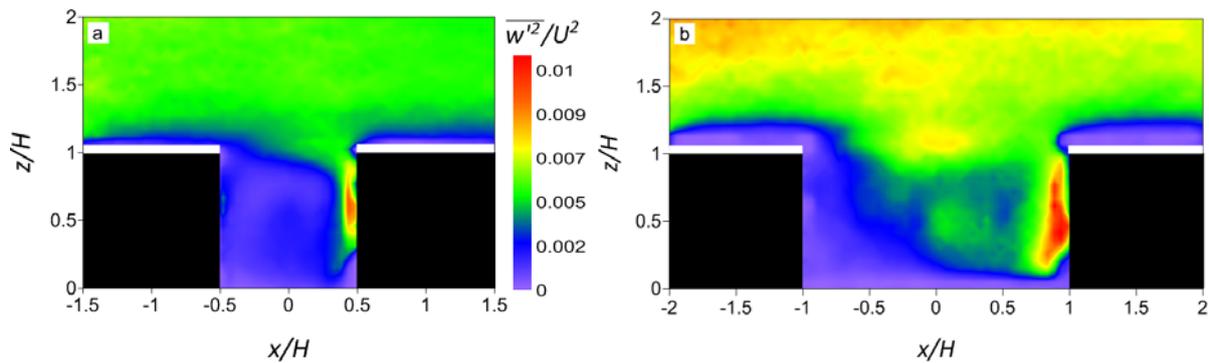

**Fig. 4** As in Fig. 3, but for the non-dimensional vertical velocity variance $\overline{w'^2}/U^2$

### 3.2 Reynolds stress

Maps of the non-dimensional, vertical momentum flux, $\overline{u'w'}/U^2$, for $AR = 1$ and 2 are depicted in Fig. 5; $\overline{u'w'}/U^2$ is negative outside the canyon for both aspect ratios, in agreement with results found in the literature (see, for example, Kastner-Klein and Rotach 2004). In contrast, inside the canyon $\overline{u'w'}/U^2$ depends strongly on $AR$. For $AR = 1$ (Fig. 5a) $\overline{u'w'}/U^2$ is positive for $\frac{z}{H} \lesssim 0.8$, except in the region close to the leeward building wall. For $AR = 2$ (Fig. 5b), $\overline{u'w'}/U^2$ within the canyon differs substantially from that observed for $AR = 1$ since it is negative everywhere except for some large, positive values within the right-half portion of the canyon. For $AR = 2$, on the other hand, outside the canyon $\overline{u'w'}/U^2$ shows inhomogeneities along $x$, a well-defined region of negative values with a maximum located at ($x / H \cong 0$, $\frac{z}{H} \cong 1.25$) and a region of positive values above the building rooftops. Inhomogeneities observed in the outer layer for $AR = 2$ conform to the different nature of the wake-interference flow compared to the skimming flow.

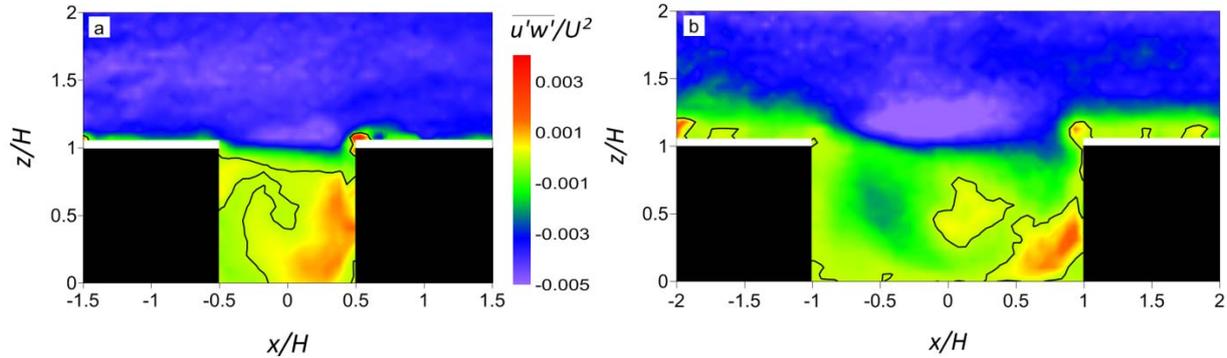

**Fig. 5** Non-dimensional vertical momentum flux $\left(\overline{u'w'}/U^2\right)$ maps for $AR = 1$ (a) and $AR = 2$ (b). The black line indicates the change in sign of $\overline{u'w'}/U^2$

A question therefore arises regarding the influence of $AR$ on the vertical structure of the outer layer. It is generally accepted that the RSL lies between the mean building height ($H$) and $z = aH$, where $a \approx 2$ (or even less) for regular structures of the urban fabric (Rotach, 1999). Above the RSL the ISL exists, where the turbulent fluxes are nearly independent of height and the streamwise velocity assumes the canonical logarithmic law. This fact is shown in Fig. 6, where the vertical profiles of the streamwise velocity component, $\langle \overline{u}(z/H) \rangle$ and turbulent stress $\langle \overline{u'w'}(z/H) \rangle$ are given for $AR = 1$, 2 and for two additional aspect ratios, namely $AR = 1.5$ and 1.75.

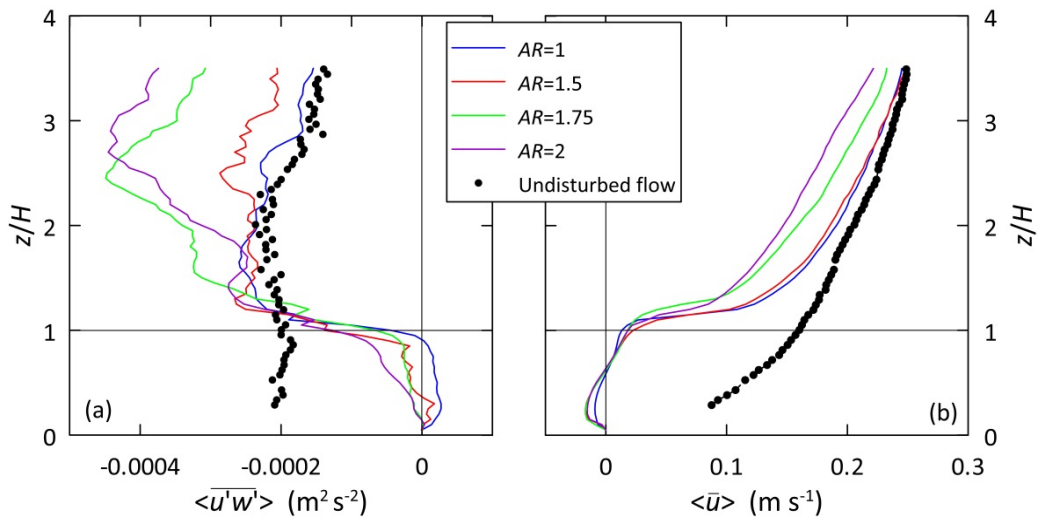

**Fig. 6** (a) Vertical profiles of the vertical momentum flux $\langle \overline{u'w'}(z/H) \rangle$ averaged along the $x$-axis for different aspect ratios $AR$. (b) as in (a), but for the streamwise velocity component $\langle \overline{u}(z/H) \rangle$

Here, we denote <·> as the spatial averaging performed along the $x$-axis in the area overlaying the canyon top and one rooftop. It is apparent that the outer flow depends strongly on the kind of flow regime. For $AR$ = 1 and 1.5 (skimming flow) the $\left\langle \overline{u'w'}(z/H) \right\rangle$ maximum occurs at $z/H \approx 1.25$. Moreover, it remains nearly constant until $z/H \approx 2.7$ and $z/H \approx 3$ for $AR$ = 1 and $AR$ = 1.5, respectively. Therefore, $z/H \approx 1.25$ could be viewed as the upper boundary of the RSL, while a well-defined ISL is present above. Note that $\left\langle \overline{u'w'}(z/H) \right\rangle$ and $\left\langle \overline{u}(z/H) \right\rangle$ share nearly the same profile when $AR$ = 1 and $AR$ = 1.5, suggesting that in the case of a skimming flow those quantities are practically insensitive to the precise value of $AR$. In contrast, for $AR$ = 1.75 and 2 (wake interference regime) the maximum of $\left\langle \overline{u'w'}(z/H) \right\rangle$ takes place above $z/H \approx 3$ and the constant-flux layer does not seem to be present.

The corresponding streamwise velocity profiles (Fig. 6b) follow the usual rough-wall logarithmic law for $\frac{z}{H} \gtrsim 1.7$ when $AR$ = 1 and 1.5, whereas the logarithmic law does not hold for $AR$ = 1.75 and 2. This implies that the ISL is not present for the wake-interference regime. If one looks at the vertical profiles of $\left\langle \overline{u}(z/H) \right\rangle$ and $\left\langle \overline{u'w'}(z/H) \right\rangle$ measured for the undisturbed flow (dotted lines in the figures) one might affirm that the ISL is eroded from below by the RSL, or, according to the analysis of Rotach (1999), the RSL completely fills-up the undisturbed surface layer existing upwind the urban canopy.

It is worthwhile noting that for $AR$ = 1.75 and 2, given the large vertical variations of $\left\langle \overline{u'w'}(z/H) \right\rangle$ and the simultaneous absence of a logarithmic law of the streamwise velocity component, the Monin-Obukhov similarity theory is difficult (or even impossible) to apply. In particular, it is not clear how to set a suitable value for the friction velocity, $u_*$, used in the classical velocity law:

$$\frac{\overline{u}(z)}{u_*} = \frac{1}{k} \ln\left( \frac{z - d_0}{z_0} \right) \tag{1}$$

where $d_0$ is the displacement height, $z_0$ is the aerodynamic roughness length and $k$ = 0.4 is the von Karman constant. In contrast, for $AR$ = 1 and 1.5 the friction velocity deduced from $\left\langle \overline{u'w'}(z/H) \right\rangle$ averaged in the ISL is $u_* \approx 0.0153$ m s$^{-1}$. Using this value as the slope to fit $\overline{u}(z)$ in the ISL to Eq. 1 one obtains $z_0$ = 0.00014 m and $d_0$ = 0.0182 m. The latter value nearly conforms to $d_0$ = 0.8 $H$ usually adopted in literature as well as to the height corresponding to the change of sign of the vertical momentum flux within the canyon (see Fig. 5a).

Finally, since the determination of $d_0$ is sometimes based on the integration along $z$ of the vertical profile of $\overline{u'w'}$ within the canyon (Jackson 1981), the presence of the large spatial inhomogeneity described above suggests a certain degree of caution is required when using such integral methods.

### 3.3 Skewness factors
Knowledge of skewness factors can be useful, e.g. to dispersion modellers, in that they are included in particle trajectory equations of Lagrangian stochastic models (see, for example, Monti and Leuzzi (1996) and references cited therein). Figures 7 and 8 show, respectively, maps of the skewness factors of the horizontal, $Sk_u = \overline{u'^3}\big/\left(\overline{u'^2}\right)^{\frac{3}{2}}$, and vertical, $Sk_w = \overline{w'^3}\big/\left(\overline{w'^2}\right)^{\frac{3}{2}}$, velocity components for $AR = 1$ and 2.

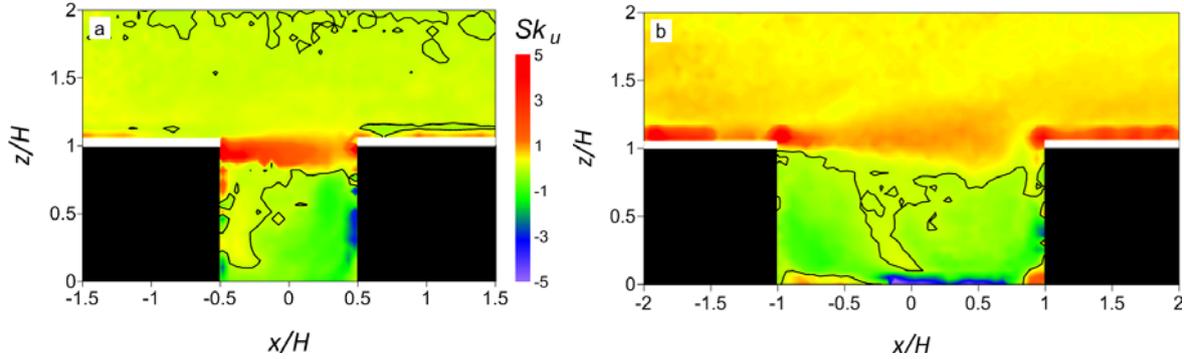

**Fig. 7** Horizontal velocity skewness factor $Sk_u$ maps for $AR = 1$ (a) and $AR = 2$ (b). The *black line* identifies the transition from negative to positive values

Overall, the (absolute) horizontal velocity skewness factor is greater than the vertical one, while it tends to assume small values for $z/H > 1.5$. As with the variance, changes of the horizontal component are small irrespective of $AR$ values inside the canyon. In particular, $Sk_u$ is negative almost everywhere for both $AR$ values, except near the canyon top, where a region of large, positive $Sk_u$ occurs. For $AR = 2$, large (positive) $Sk_u$ is located also near the building tops (Fig. 7b). In contrast, the sign of $Sk_w$ inside the canyon conforms to that of $\overline{w}$ (see Fig. 2), i.e. negative close to the windward building wall and positive within the left-half part of the canyon.

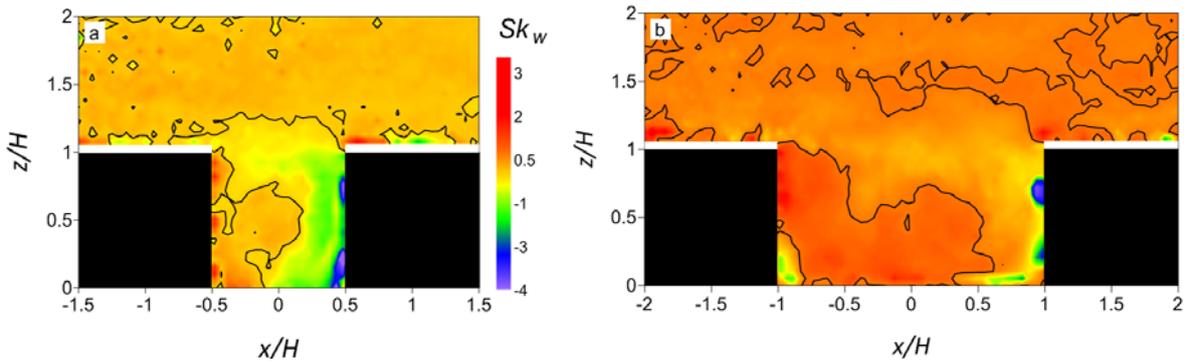

**Fig. 8** As in Fig. 7, but for the vertical velocity skewness factor $Sk_w$

### 3.4 Shear production and viscous dissipation rate of the turbulent kinetic energy

Here, we focus on the turbulent kinetic energy ($\overline{q}$) budget equation, in particular on the shear production term, $P = -\overline{v_i'v_j'}\frac{\partial \overline{v_i}}{\partial x_j}$, and on the rate of dissipation of $\overline{q}$, namely $\varepsilon = 2\nu \overline{\left(\partial v_i'/\partial x_j'\right)^2}$, where $i = 1,2,3$ and $j = 1,2,3$ indicate the axis of the coordinate system, while $v_i$ is the velocity component along the i-axis. Both $P$ and $\varepsilon$ offer important insights into the nature of the turbulence. $P$ represents a loss for the mean kinetic energy and a gain for the turbulence and it is expected to be positive in shear flows. In our case, given the two-dimensional nature of the flow, $P$ reduces to,

$$P = -\overline{u'^2}\frac{\partial \overline{u}}{\partial x} - \overline{u'w'}\frac{\partial \overline{u}}{\partial z} - \overline{w'u'}\frac{\partial \overline{w}}{\partial z} - \overline{w'^2}\frac{\partial \overline{w}}{\partial z} = P_{uu} + P_{uw} + P_{wu} + P_{ww} \qquad (2)$$

The four terms in Eq. 2 are normalized by $U^3/H$ and depicted separately in Figs. 9, 10, 11 and 12.

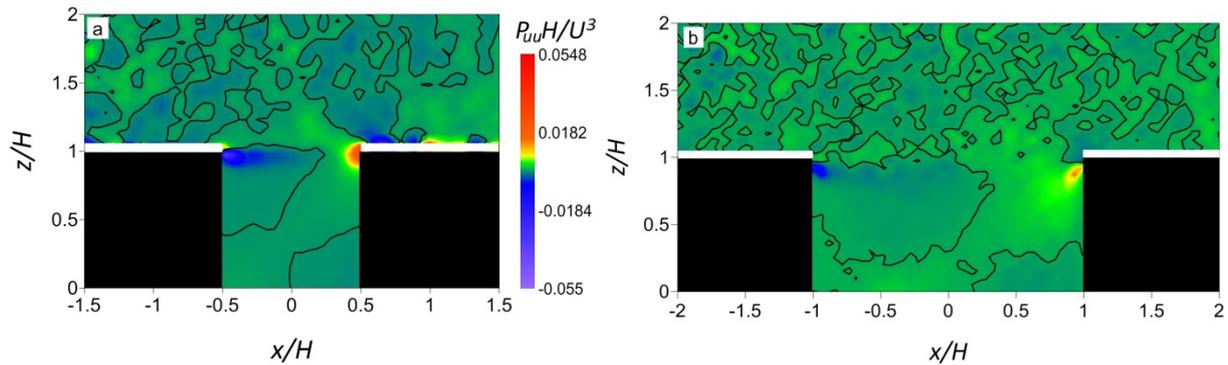

**Fig. 9** $P_{uu} H/U^3$ for $AR$ = 1 (a) and $AR$ = 2 (b). The *black line* identifies the transition from negative to positive values

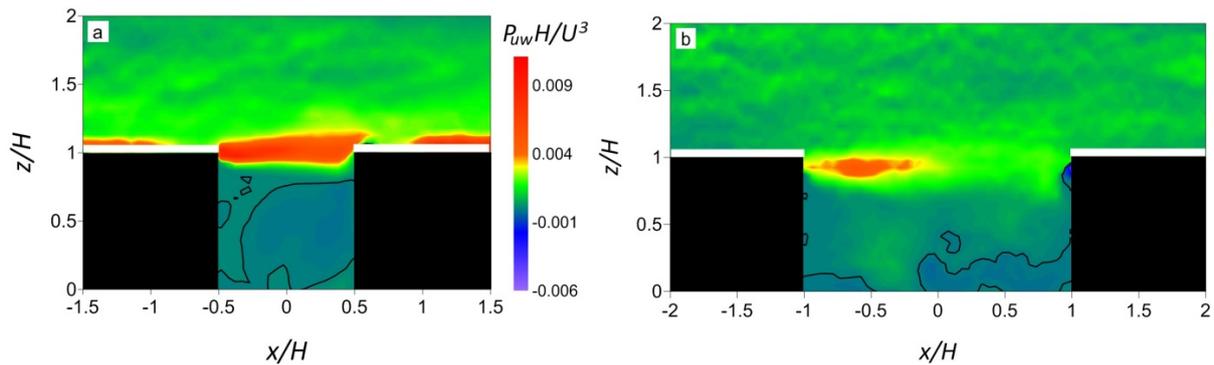

**Fig. 10** As in Fig. 9, but for $P_{uw} H/U^3$

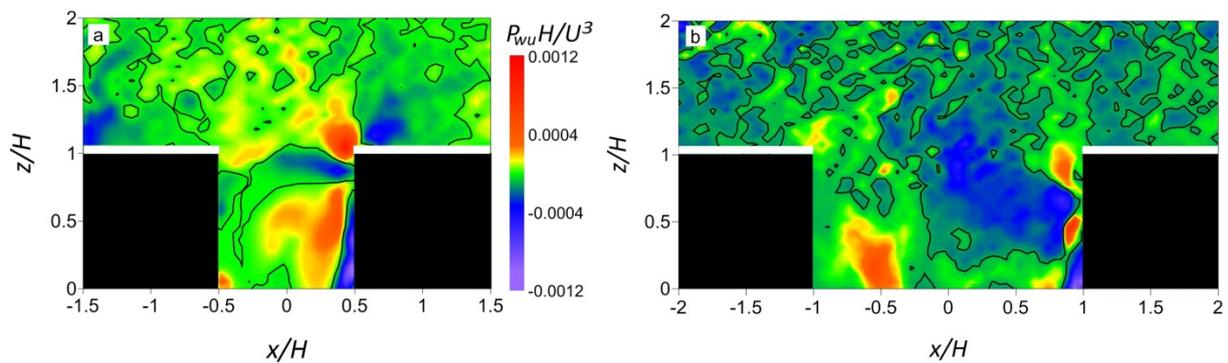

**Fig. 11** As in Fig. 9, but for $P_{wu} H/U^3$

The large negative (positive) values of $P_{uu} H/U^3$ occurring at the canyon top for both $AR$ values are related to the increase (decrease) along $x$ of the streamwise velocity component associated with separation from (re-attachment to) the building rooftop (Figs. 9a and b). Similarly, within the canyon positive (negative) values of $P_{uu} H/U^3$ correspond to regions of decrease (increase) of $\overline{u}$ along the $x$-axis. The changes in sign shown by $P_{uu} H/U^3$ above the rooftops are, probably, a result of the small variations of $\overline{u}$ along $x$, accentuated by large values of $\overline{u'^2}$ occurring therein. Given the uncertainties in the performance of the acquisition procedure above the buildings, those values must be considered carefully. Large (positive) values of $P_{uw} H/U^3$ (Figs. 10a and b) occur near the canyon top for both $AR$ values and, in general, within the RSL. For $AR$ = 1, since $\overline{u'w'}$ is mostly negative within the canyon, the sign of $P_{uw} H/U^3$ is mainly related to that assumed by $\partial\overline{u}/\partial z$ therein. Similar considerations hold for $P_{wu} H/U^3$ (Fig. 11), even though the large positive (negative) values of $\partial\overline{w}/\partial z$ close to the facing wall of the windward buildings give rise to large negative (positive) $P_{wu} H/U^3$. Similarly to $P_{uu} H/U^3$, the sign of $P_{ww} H/U^3$ (Fig. 12) depends only on that of the velocity gradient. Therefore, it is nearly zero above the RSL and reaches local maxima within the canopy as a result of local variations in $\partial\overline{w}/\partial z$.

Finally, the non-dimensional production term $PH/U^3$ is positive above the canyon top particularly for $AR$ = 1 (Fig. 13a), where a well-defined region of maxima is present. That area corresponds with the mixing layer that develops after the trailing edge of the upstream obstacle, which is characterized by a strong vertical shear. Similar results were obtained by Salizzoni et al. (2011). The positive and negative peaks occurring over the rooftops are related to those shown by $P_{uu} H/U^3$. However, as mentioned above, the results obtained in those regions must be viewed with circumspection. For $AR$ = 2 (Fig. 13b) the region of large $PH/U^3$ is still present, even though it is less evident with respect to $AR$ = 1.

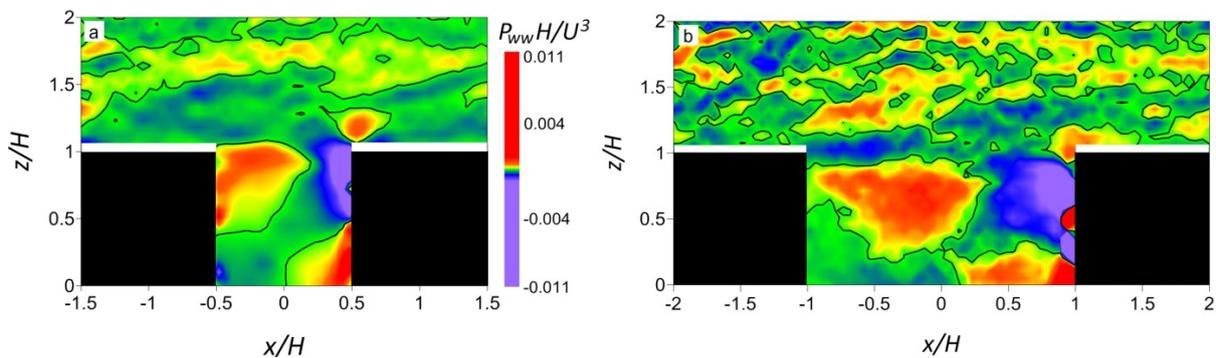

**Fig. 12** As in Fig. 9, but for $P_{ww} H/U^3$

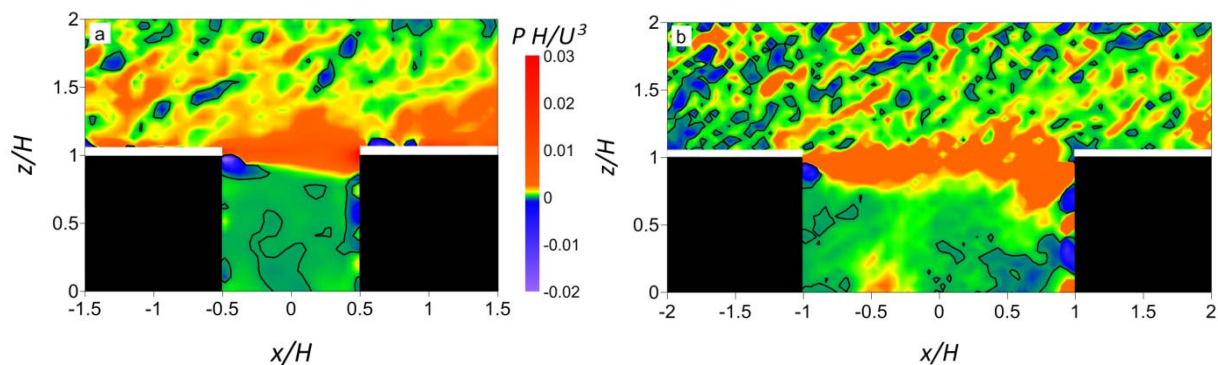

**Fig. 13** As in Fig. 9, but for $PH/U^3$

As is well-known, the determination of the dissipation rate $\varepsilon$ consists of several terms, of $\overline{\left(\partial v_i'/\partial x_j\right)^2}$. In our case, given the lack of information about the velocity components along the $y$-axis, only an indirect estimation of $\varepsilon$ can be performed. Several simplified expressions for $\varepsilon$ are available in the literature, such as analytical formulations (Sawford 2006), parametrizations derived from similarity theories (Cassiani et al. 2005) and other estimations based on mean velocity gradients (Stull 1988). In particular, here the approximation valid for isotropic turbulence was used (Hinze 1975, hereinafter referred to as *estimate I*), viz.,

$$\varepsilon_I = \frac{15}{4}\nu\left[\overline{\left(\frac{\partial u'}{\partial z}\right)^2}+\overline{\left(\frac{\partial w'}{\partial z}\right)^2}\right] \qquad (3)$$

This formulation is particularly suitable for the present study in that the dissipation rate can be calculated using only the two components of the strain rate tensor obtained through the measured vertical profiles of the velocity vector. A simpler method (*estimate II*) is based on the knowledge of the mean strain rate tensor (Stull 1988):

$$\varepsilon_{II} = 0.3\overline{q}\sqrt{\left(\frac{\partial \overline{u}}{\partial x}+\frac{\partial \overline{u}}{\partial z}+\frac{\partial \overline{w}}{\partial x}+\frac{\partial \overline{w}}{\partial z}\right)^2} \qquad (4)$$

The turbulent kinetic energy in Eq. 4 was approximated by $\overline{q} = \left[2\overline{u'^2}+\overline{w'^2}\right]\Big/2$. Maps reporting both the estimations of $\varepsilon$ (normalized by the factor $U^3/H$) for $AR$ = 1 are depicted in Fig. 14. Quite surprisingly, they agree reasonably well both inside and outside the canyon. Lower values occur within the canyon, particularly near the leeward building, in consonance with the pattern shown by the $\overline{q}$ (not shown). Above the canyon, $\varepsilon$ reaches higher values, with maxima above the rooftops. These peaks are higher for $\varepsilon_I H/U^3$ (Fig. 14a), even though the approximations introduced above suggest considering the results with a certain degree of caution.

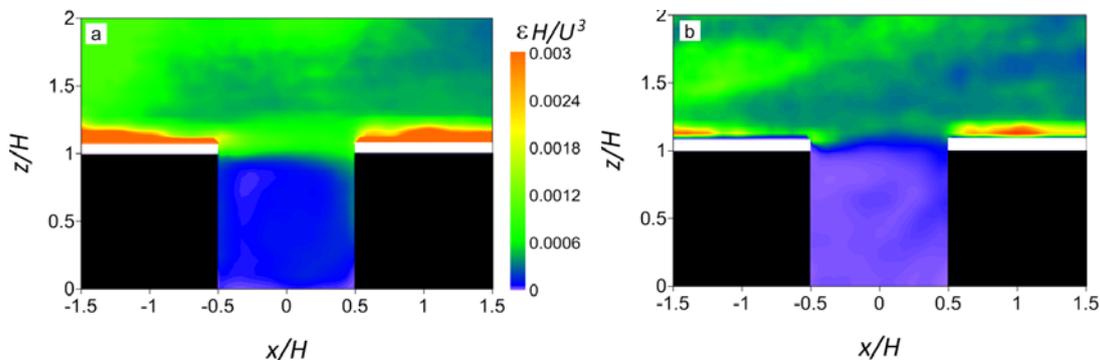

**Fig. 14** Maps of $\varepsilon_I H/U^3$ (a) and $\varepsilon_{II} H/U^3$ (b) for $AR$ = 1.

**4 Conclusions**

The mean flow and turbulence were studied within and above an idealized, 2D urban canopy layer using a water-channel facility. The feature tracking technique was used to acquire velocity data in a vertical plane parallel to the streamwise direction. Specifically, 2D maps of mean velocities, velocity variances, vertical momentum flux and skewness were presented for the aspect ratios $AR = 1$ and 2. Attention was also focussed on the analysis of various terms in the $TKE$ $(\overline{q})$ budget equation and the dissipation rate $\varepsilon$.

For $AR = 1$ (skimming flow regime), the mean and the variance of both the velocity components agree reasonably well with those reported in the literature. For $AR = 2$ (wake-interference regime), a clear deviation of those parameters is observed within the canyon with respect to the $AR = 1$ case; for the outer flow the dependence on $AR$ seems to be of second order. In contrast, the vertical momentum flux depends strongly on $AR$ both inside and outside the canyon. For $AR = 1$ the vertical momentum flux shows a quasi-constant layer, i.e. the ISL, for $1.25 \leq \frac{z}{H} \leq 2.7$ , while for $AR = 2$ the ISL seems to be absent. This fact makes the application of the Monin-Obukhov similarity theory dubious. Within the canyon, the vertical momentum flux is generally positive for $AR = 1$ and negative for $AR = 2$, except in a small region located within the right-half of the canyon. This could give rise to problems in the application of consolidated methods used to calculate the displacement height in the case of the wake-interference regimes. The dissipation rate $\varepsilon$ is calculated by using two different approaches: the first one is based on the estimation of two components of the fluctuating strain rate tensor, the second method requires (only) the knowledge of the mean strain rate tensor. Both determinations give similar results, and this suggests that the latter, simpler expression of $\varepsilon$ might be used with a certain degree of reliability.